\documentclass[5p,longtitle]{elsarticle}

\usepackage{lineno,hyperref}
\usepackage{color}
\usepackage{natbib}
\usepackage{amsmath}
\usepackage[normalem]{ulem}
\usepackage{soul}

\newcommand{\qsqU}{$(\text{GeV}/c)^2$}
\newcommand{\di}[0]{\mathrm{d}}

\usepackage{mathtools}

\usepackage{bm}
\usepackage{amsfonts}
\usepackage{mathrsfs}
\usepackage{amssymb}

\usepackage{booktabs}

\usepackage[dvipsnames]{xcolor}

\newcommand{\xBj}[0]{x_{\text{Bj}}}

\newcommand{\PluLeft}{{\stackrel{\textstyle{+}}{\textstyle{\leftarrow}}}}
\newcommand{\MinRight}{{\stackrel{\textstyle{-}}{\textstyle{\rightarrow}}}}
\newcommand{\LeftRight}{\leftrightarrows}
\newcommand{\phigg}{\phi}

\journal{Physics Letters B}

\begin{document}

\begin{frontmatter}

\title{
Transverse extension of partons in the proton \\ 
probed in the sea-quark range by measuring the DVCS cross section\\
[15pt] {\normalsize COMPASS Collaboration}}

\author[dubna]{R.~Akhunzyanov}
\author[turin_u]{M.~G.~Alexeev}
\author[dubna]{G.~D.~Alexeev}
\author[turin_u,turin_i]{A.~Amoroso}
\author[illinois,saclay]{V.~Andrieux}
\author[dubna]{N.~V.~Anfimov}
\author[dubna]{V.~Anosov}
\author[dubna]{A.~Antoshkin}
\author[dubna,praguectu]{K.~Augsten}
\author[warsaw]{W.~Augustyniak}
% ...
\address[dubna]{Joint Institute for Nuclear Research, 141980 Dubna, Moscow region, Russia\fnref{E}}
% ...
\address[praguectu]{Czech Technical University in Prague, 16636 Prague, Czech Republic\fnref{V}}
% ...
\address[saclay]{IRFU, CEA, Universit\'e Paris-Saclay, 91191 Gif-sur-Yvette, France\fnref{U}}
% ...
\address[triest_u]{University of Trieste, Dept.\ of Physics, 34127 Trieste, Italy}
\address[triest_i]{Trieste Section of INFN, 34127 Trieste, Italy}
\address[turin_u]{University of Turin, Dept.\ of Physics, 10125 Turin, Italy}
\address[turin_i]{Torino Section of INFN, 10125 Turin, Italy}
% ...
\address[illinois]{University of Illinois at Urbana-Champaign, Dept.\ of Physics, Urbana, IL 61801-3080, USA\fnref{c}}
\address[warsaw]{National Centre for Nuclear Research, 00-681 Warsaw, Poland\fnref{d}}
% ...
\fntext[A]{Also at Instituto Superior T\'ecnico, Universidade de Lisboa, Lisbon, Portugal}
\fntext[B]{Also at Dept.\ of Physics, Pusan National University, Busan 609-735, Republic of Korea and at Physics Dept., Brookhaven National Laboratory, Upton, NY 11973, USA}
\fntext[C]{Also at Abdus Salam ICTP, 34151 Trieste, Italy}
\fntext[D]{Supported by the DFG cluster of excellence `Origin and Structure of the Universe' (www.universe-cluster.de) (Germany)}
\fntext[E]{Supported by CERN-RFBR Grant 12-02-91500}
\fntext[F]{Supported by the Laboratoire d'excellence P2IO (France)}
\fntext[G]{Present address:  University of Connecticut, Storrs, Connecticut 06269, US}
\fntext[H]{Supported by the DFG Research Training Group Programmes 1102 and 2044 (Germany)}
\fntext[I]{Also at Chubu University, Kasugai, Aichi 487-8501, Japan\fnref{Y}}
\fntext[J]{Also at Dept.\ of Physics, National Central University, 300 Jhongda Road, Jhongli 32001, Taiwan}
\fntext[K]{Also at KEK, 1-1 Oho, Tsukuba, Ibaraki 305-0801, Japan}
\fntext[L]{Present address: Universit\"at Bonn, Physikalisches Institut, 53115 Bonn, Germany}
\fntext[M]{Also at Moscow Institute of Physics and Technology, Moscow Region, 141700, Russia}
\fntext[N]{Also at Yerevan Physics Institute, Alikhanian Br. Street, Yerevan, Armenia, 0036}
\fntext[O]{Also at Dept.\ of Physics, National Kaohsiung Normal University, Kaohsiung County 824, Taiwan}
\fntext[P]{Also at Institut f\"ur Theoretische Physik, Universit\"at T\"ubingen, 72076 T\"ubingen, Germany}
\fntext[Q]{Also at University of Eastern Piedmont, 15100 Alessandria, Italy}
\fntext[R]{Present address: RWTH Aachen University, III.\ Physikalisches Institut, 52056 Aachen, Germany}
\fntext[S]{Present address: Uppsala University, Box 516, 75120 Uppsala, Sweden}
\fntext[T]{Supported by BMBF - Bundesministerium f\"ur Bildung und Forschung (Germany)}
\fntext[U]{Supported by FP7, HadronPhysics3, Grant 283286 (European Union)}
\fntext[V]{Supported by MEYS, Grant LG13031 (Czech Republic)}
\fntext[W]{Supported by B.Sen fund (India)}
\fntext[X]{\raggedright   Supported by FCT - Funda\c{c}\~{a}o para a Ci\^{e}ncia e Tecnologia, COMPETE and QREN, Grants CERN/FP 116376/2010, 123600/2011  and CERN/FIS-NUC/0017/2015 (Portugal)}
\fntext[Y]{Supported by MEXT and JSPS, Grants 18002006, 20540299, 18540281 and 26247032, the Daiko and Yamada Foundations (Japan)}
\fntext[Z]{Supported by the Ministry of Science and Technology (Taiwan)}
\fntext[a]{Supported by the Israel Academy of Sciences and Humanities (Israel)}
\fntext[b]{Supported by the Russian Federation  program ``Nauka'' (Contract No. 0.1764.GZB.2017) (Russia)}
\fntext[c]{Supported by the National Science Foundation, Grant no. PHY-1506416 (USA)}
\fntext[d]{Supported by NCN, Grant 2017/26/M/ST2/00498 (Poland)}
\fntext[*]{Deceased}
\cortext[cors]{Corresponding authors}
\begin{abstract}
We report on the first measurement of exclusive single-photon muoproduction on the proton by COMPASS using 160\,GeV/$c$ polarised $\mu^+$ and $\mu^-$ beams of the CERN SPS impinging on a liquid hydrogen target. We determine the dependence of the average of the measured $\mu^+$ and $\mu^-$ cross sections for deeply virtual Compton scattering on the squared four-momentum transfer $t$ from the initial to the final proton. The slope $B$ of the $t$-dependence is fitted with a single exponential function, which yields 
$B=(4.3 \ \pm \ 0.6_{\text{stat}}\ _{- \ 0.3}^{+ \ 0.1}\big\rvert_{\text{sys}})\,(\text{GeV}/c)^{-2}$.
This result can be converted into a transverse extension of partons in the proton, 
{$\sqrt{\langle r_{\perp}^2 \rangle} = (0.58 \ \pm 0.04_{\text{stat}} \ _{- \ 0.02}^{+ \ 0.01}\big\rvert_{\text{sys}}\ 
\pm 0.04_{\text{model}})\, \text{fm}$}. For this measurement, the average virtuality of the photon mediating the interaction is $\langle Q^2 \rangle = 1.8\,(\text{GeV/}c)^2$ and the average value of the Bjorken variable is $\langle \xBj \rangle = 0.056$. 
\end{abstract}


\end{frontmatter}

\section{Introduction}
The structure of the proton has been studied over 
half a century, still its understanding constitutes one of the very important challenges that physics is facing today. Quantum Chromodynamics (QCD), the theory of strong interaction that governs the dynamics of quarks and gluons as constituents of the proton, is presently not analytically solvable. Lepton-proton scattering experiments have been proven to be very powerful tools to unravel the internal dynamics of the proton: (i) elastic scattering allows access to charge and current distributions in the proton by measuring electromagnetic form factors; (ii) deep-inelastic scattering (DIS) provides important information 
on the density distributions as a function of 
longitudinal momentum for quarks and gluons in the proton, 
encoded in universal parton distribution functions. 

Deeply virtual Compton scattering (DVCS), $\gamma^* p \rightarrow \gamma p$, is the production of a single real photon $\gamma$ through the absorption of a virtual photon $\gamma^*$ by a proton $p$. This process combines features of the elastic process and those of the inelastic processes. Using 
the concept of generalized parton distributions (GPDs) \cite{Mueller:1998fv,Ji:1996ek, Ji:1996nm, Radyushkin:1996nd, Radyushkin:1997ki}, 
it was shown \cite{Burkardt:2000za,Burkardt:2002hr,Diehl:2002he,Ralston:2001xs} that in a certain kinematic domain DVCS allows access to correlations between transverse-position and longitudinal-momentum distributions of the partons in the proton. Here, longitudinal and transverse refer to the direction of motion of the initial proton facing the virtual photon. 
The measurement of DVCS probes the transverse extension of the parton density in the proton over the experimentally accessible region of longitudinal momentum of the active parton.
Exploring the interplay between longitudinal and transverse partonic degrees of freedom by DVCS is often referred to as ``proton tomography''.
The DVCS process is studied through exclusive single-photon production in lepton-proton scattering.
The experimental results obtained so far are discussed in a recent review~\cite{dHose:2016mda}.

In this Letter, we present the result 
on a measurement of the DVCS cross section obtained by
studying exclusive single-photon production in muon-proton scattering, $\mu p \rightarrow \mu' p' \gamma$.
Following Refs.~\cite{Burkardt:2000za,Burkardt:2002hr,Frankfurt:2005mc,Aschenauer:2013hhw,Moutarde:2018kwr}, the slope $B$
of the measured exponential $t$-dependence of the differential DVCS cross section can
approximately be converted into the average squared transverse
extension of partons in the proton as probed by DVCS,
\begin{equation}
\langle r_{\perp}^2 (\xBj) \rangle \approx 2 \langle B (\xBj) \rangle \hbar^2,
\label{eq:radius}
\end{equation}
which is measured at the average value of $\xBj$ accessed by COMPASS. The approximation 
used above is discussed in Sec.~\ref{sec:interpretation}. 
In the following  we refer to $\sqrt{\langle r_{\perp}^2 \rangle}$ as transverse
extension of partons.
Here, $t$ is the squared four-momentum transferred to the target proton, $\xBj = Q^2/(2 M \nu)$
the Bjorken variable, $Q^2=-(k_{\mu}-k_{\mu'})^2$, and $\nu=(k_{\mu}^0-k_{\mu'}^0)$ the energy of the virtual photon in the target rest frame, with $k_{\mu}$
and $k_{\mu'}$ denoting the four-momenta of the incoming and scattered muon, respectively, and $M$ the proton mass.
The quantity $r_\perp$ is the transverse distance between the active quark and the center of momentum
of the spectator quarks and is hence used in this Letter to represent the transverse extension of partons in the proton.

\begin{figure}[h!] 
\begin{center}  
  \includegraphics[width=0.9\linewidth]{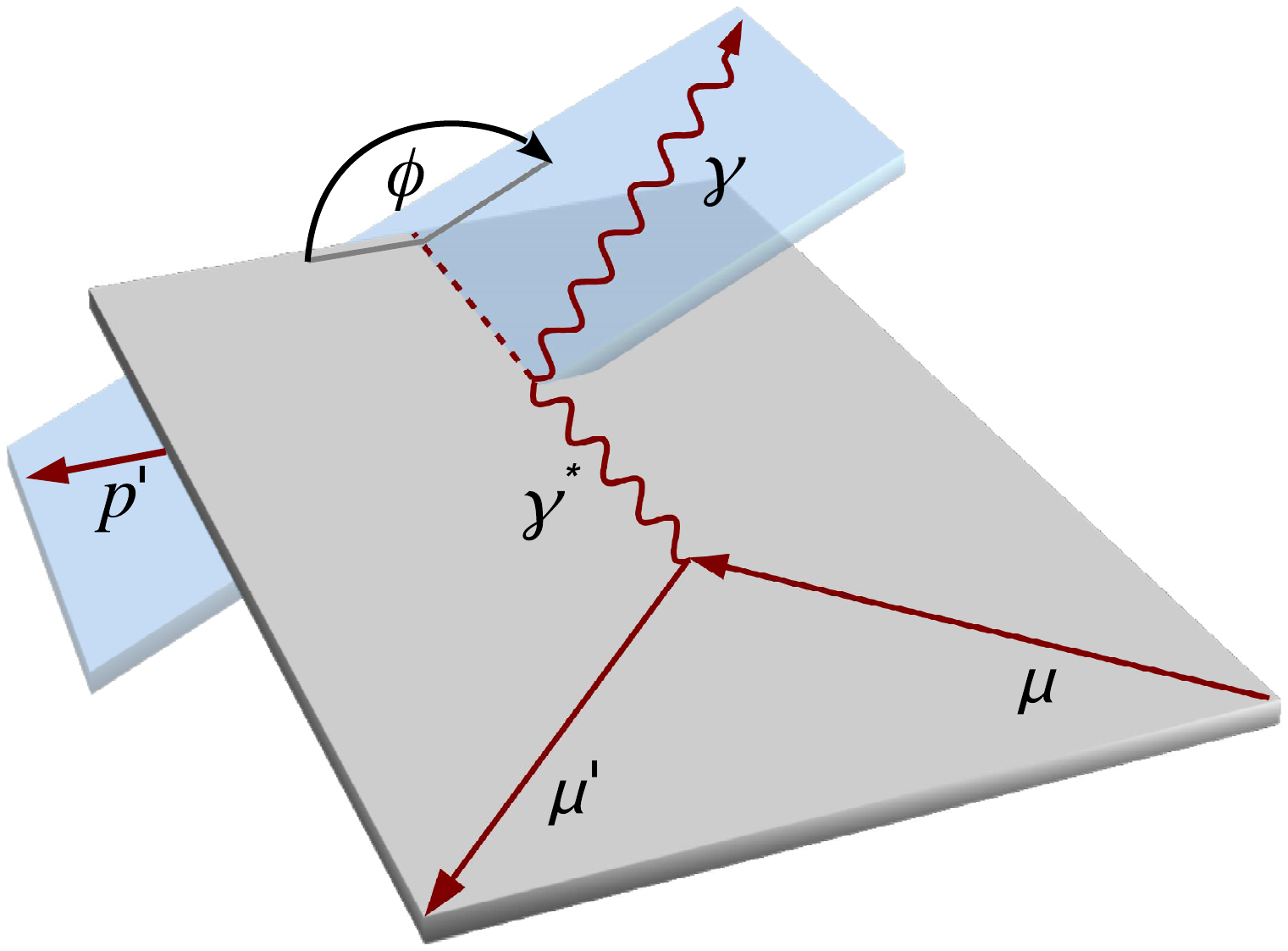}
\end{center} 
\vspace{-0.25cm}
  \caption{\label{fig:phi_angle}
  Definition of $\phigg$, the azimuthal angle between the lepton-scattering and photon-production planes.}
\end{figure}

Using boldface letters for particle three-momenta, defining $\textbf{q}=\textbf{k}_{\mu}-\textbf{k}_{\mu'}$, denoting
by $\textbf{p}_{\gamma}$ the momentum of the real photon, and calculating the azimuthal angle between the lepton-scattering and photon-production planes (see also Fig.~\ref{fig:phi_angle}) as
\begin{equation}
\phigg=\frac{(\textbf{q} \times \textbf{k}_{\mu})\cdot \textbf{p}_{\gamma}}{|(\textbf{q} \times \textbf{k}_{\mu})\cdot \textbf{p}_{\gamma}|}
\text{arccos}\Biggl(\frac{(\textbf{q} \times \textbf{k}_{\mu}) \cdot (\textbf{q} \times \textbf{p}_{\gamma})}{|\textbf{q} \times \textbf{k}_{\mu}| |\textbf{q} \times \textbf{p}_{\gamma}|}\Biggr),
\end{equation}
the cross section of muon-induced single-photon production is written as
\begin{equation} 
\di \sigma:=\frac{\di^4 \sigma^{\mu p }}{\di Q^2 \di \nu \di t \di \phigg}.
\end{equation}
This cross section was measured separately using either a $\mu^+$ or a $\mu^-$ beam of 160\,GeV/$c$ average momentum, which was provided by the M2 beamline of the CERN SPS. The natural polarisation of the muon beam originates from the parity-violating decay-in-flight of the parent mesons, which implies opposite signs of the polarisation for the used $\mu^+$ and $\mu^-$ beams. For both beams, the absolute value of the average beam polarisation is about 0.8 with an uncertainty of about 0.04.
Denoting charge and helicity of an incident muon by $\pm$ and $\LeftRight$, respectively, the sum of the cross sections for $\mu^+$ and $\mu^-$ beams reads:
\begin{equation} 
2\, \di \sigma \equiv \di\sigma^{\PluLeft} + \di\sigma^{\MinRight}
= 2 (\di\sigma^{BH} + \di\sigma^{DVCS} - |P_\mu| \di\sigma^{I}).
\label{eq:ssc}
\end{equation}
Here, $P_\mu$ denotes the polarisation of the muon beam. The single-photon final state in lepton-nucleon scattering can also originate from the Bethe-Heitler (BH) process, i.e. photon emission from either the incoming or the outgoing lepton. Hence the DVCS and BH processes interfere, so that the above sum of $\mu^+$ and $\mu^-$ cross sections comprises not only the contributions $\di\sigma^{DVCS}$ and $\di\sigma^{BH}$ but also that from the interference term denoted by $\di\sigma^{I}$. 
 
At sufficiently large values of $Q^2$ and small values of $|t|$, the azimuthal dependences of the DVCS cross section and 
\textcolor{blue}{of} the interference term including twist-3 contributions read as follows~\cite{mueller2002}:
\begin{equation} 
\nonumber
\di\sigma^{DVCS} \propto \frac{1}{y^2Q^2}
(c_0^{DVCS} + c_1^{DVCS} \cos{\phigg} + c_2^{DVCS} \cos{2 \phigg)},
\end{equation}
\begin{equation}
\di\sigma^{I} \propto \frac{1}{\xBj y^3 t P_1(\phigg)P_2(\phigg)}
(s_1^{I} \sin{\phigg} + s_2^{I} \sin{2 \phigg}).
\label{eq:phi_mods}
\end{equation}
Here, $P_1(\phigg)$ and $P_2(\phigg)$ are the BH lepton propagators, $y$ is the fractional energy of the virtual photon, and $c_i^{DVCS}$ and $s_i^{I}$ are related to certain combinations of Compton Form Factors (CFFs)~\cite{mueller2002}. The latter are convolutions of GPDs with 
functions describing the Compton interaction at the parton level.
At leading order in the strong coupling constant $\alpha _{S}$ and using the leading-twist approximation, in  Eq.~(\ref{eq:phi_mods}) only the terms containing  $c_0^{DVCS}$ and $s_1^{I}$ remain. In terms of 
Compton helicity amplitudes, this corresponds to the dominance of the amplitude that describes the transition from a transversely polarized virtual photon to a transversely polarised real photon.

After subtracting the cross section of the BH process, $\di\sigma^{BH}$, from Eq.~(\ref{eq:ssc}) and integrating the remainder
over $\phigg$, all azimuth-dependent terms disappear and only the dominant contribution from transversely polarized virtual photons to the DVCS cross section remains.
It is indicated by the subscript $\text{T}$:
\begin{eqnarray} 
\frac{\di^3 \sigma^{\mu p }_{\text{T}}}
{\di Q^2 \di \nu dt}
=\int_{-\pi}^{\pi} \di \phigg \ (\di \sigma-\di\sigma^{BH}) \propto c_0^{DVCS}.
\label{eq:phi_int}
\end{eqnarray}
This cross section is converted into the cross section for virtual-photon scattering using the flux  $\Gamma(Q^2,\nu,E_{\mu})$ for transverse virtual photons, 
\begin{equation} 
\label{eq:flux_relation}
\frac{\di \sigma^{\gamma^* p }}{ \di t }= \frac{1}{\Gamma(Q^2,\nu,E_{\mu})}
\frac{\di^3 \sigma^{\mu p }_{\text{T}}}{\di Q^2 \di \nu dt},
\end{equation}
with
\begin{equation}  
\label{eq:photon_flux}
\begin{split}
\Gamma(Q^2,\nu,E_{\mu}) & = \frac{\alpha_{\text{em}}  (1- \xBj)}{2 \pi Q^2 y E_{\mu}} \Bigg[ y^2 \bigg(1 - \frac{2 m_{\mu}^2}{Q^2} \bigg) \\
& + \frac{2}{1+Q^2/\nu^2} \bigg(1-y - \frac{Q^2}{4E_{\mu}^2} \bigg) \Bigg],
\end{split}
\end{equation}
for which the Hand convention \cite{hand} is used.
Here, $m_{\mu}$ and $E_{\mu}$ denote the mass and energy of the incoming muon, respectively, and $\alpha_{\text{em}}$ the electromagnetic fine-structure constant.

\section{Experimental set-up}

The data used for this analysis were recorded during four weeks in 2012 using the COMPASS set-up. The muon beam was centered onto a {2.5 m long} liquid-hydrogen target surrounded by two concentric cylinders consisting of slats of scintillating counters, which detected recoiling protons by the time-of-flight (ToF) technique.
The first electromagnetic calorimeter (ECAL0) was placed directly downstream of the target to detect photons emitted at large polar scattering angles. Particles emitted through its central opening into the forward direction were measured using the open-field two-stage magnetic spectrometer. 
Each spectrometer stage comprised an electromagnetic calorimeter (ECAL1 or ECAL2), a hadron calorimeter, a muon filter for muon
identification, and a variety of tracking detectors. A detailed description of the spectrometer can be found in Refs.~\cite{Abbon:2007pq,Abbon:2014aex,proposal}.
The period of data taking was divided into several subperiods. After each subperiod, charge and polarisation of the muon beam were swapped simultaneously. The total integrated luminosity is 18.9\,$\text{pb}^{-1}$ for the $\mu^+$ beam with negative polarisation and 23.5\,$\text{pb}^{-1}$ for the
$\mu^-$ beam with positive polarisation. 

\section{Data analysis}

The selected events are required to have at least one reconstructed vertex inside the liquid-hydrogen target associated with an incoming muon, a single outgoing particle of the same charge, a recoil proton candidate, and exactly one ``neutral cluster'' detected above 4\,GeV, 5\,GeV or 10\,GeV in ECAL0, ECAL1, or ECAL2 respectively. Here, neutral cluster specifies a cluster not associated to a charged particle. 
For ECAL0 any cluster is considered as neutral, as there are no tracking detectors in front.
An outgoing charged particle that traverses more than 15 radiation lengths is considered to be a muon. 
The spectrometer information on incoming and scattered muons, as well as on position and energy measured for the neutral cluster, is used
together with measured information from the time-of-flight system of the target-recoil detector.
For a given event, the kinematics of all recoil proton candidates are compared with the corresponding predictions that are obtained using spectrometer information only.

\begin{figure}[h!] 
\begin{center}
\begin{tabular}{l}
\includegraphics[width=0.45\textwidth,bb=140 100 690 490,clip]{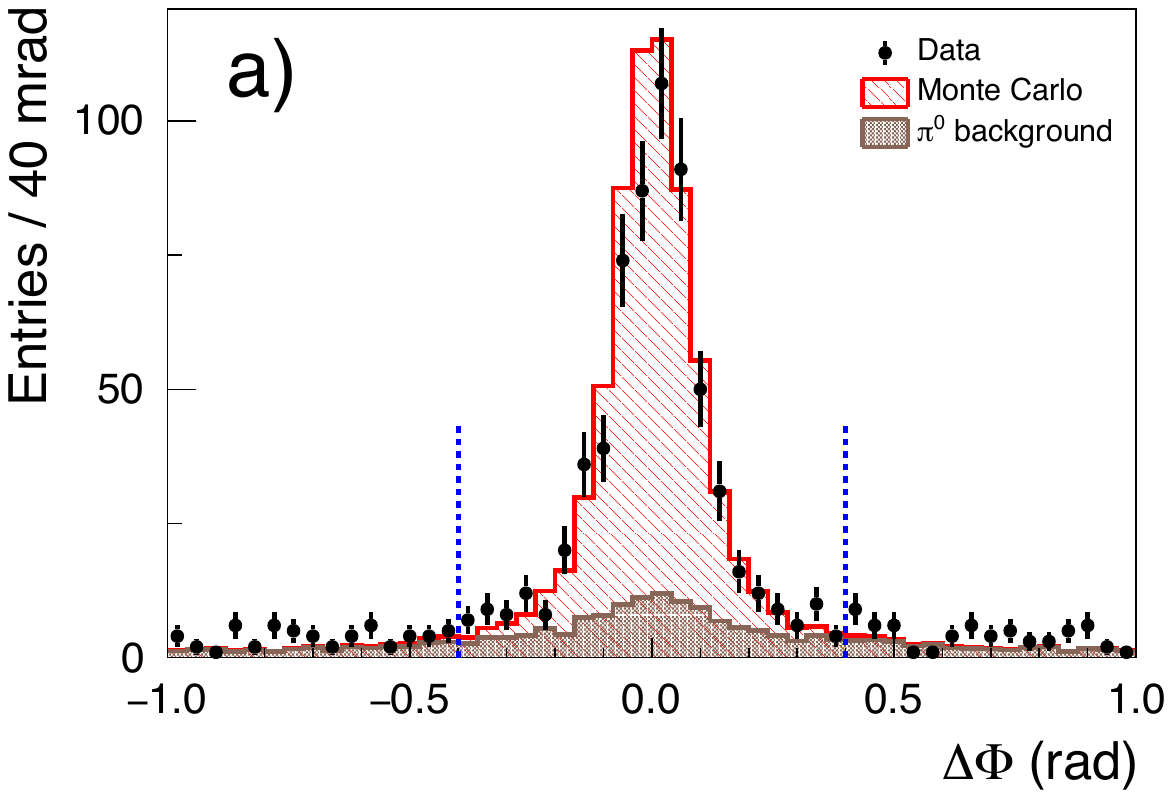}
\\
\includegraphics[width=0.45\textwidth,bb=140 100 690 490,clip]{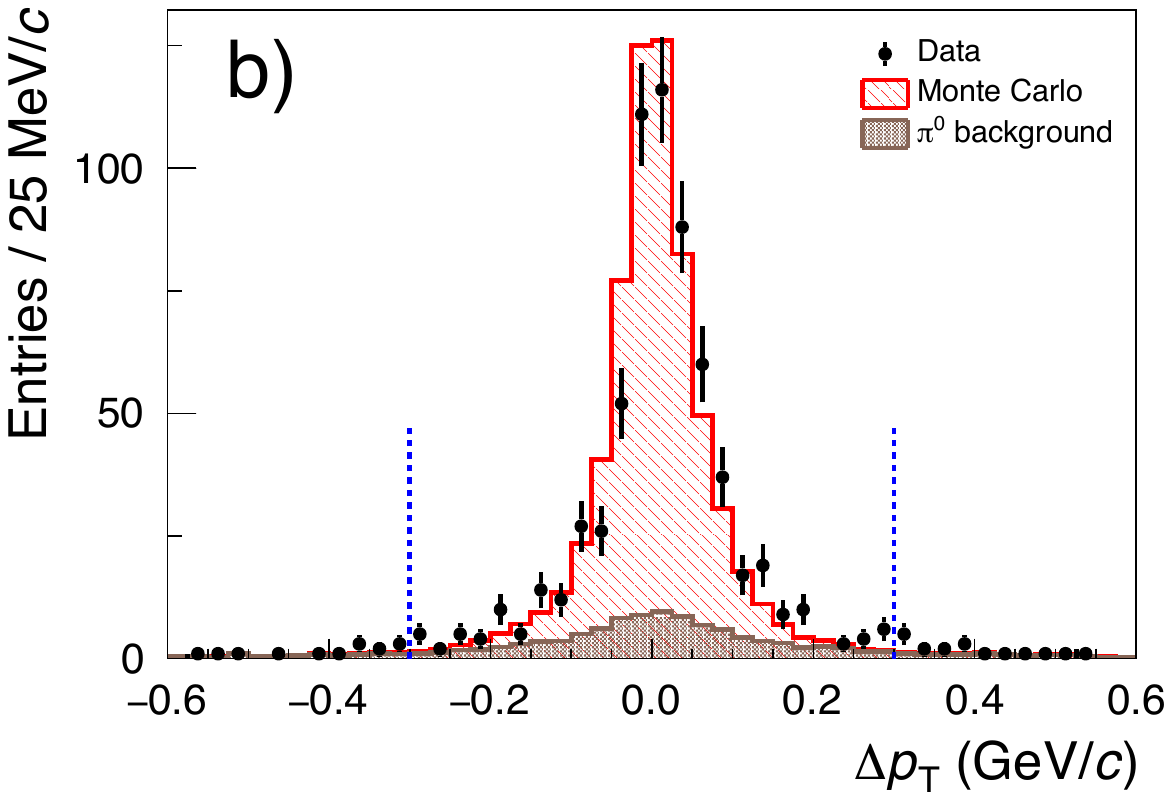}
\end{tabular}
\end{center}
\caption
{\label{fig:ex_vars} 
Distribution of the difference between predicted and reconstructed values of (a) the azimuthal angle and (b) the transverse momentum of the recoiling proton candidates for $1\,(\text{GeV/}c)^2<Q^2<5\,(\text{GeV/}c)^2$, $0.08\,(\text{GeV/}c)^2<|t|<0.64\,(\text{GeV/}c)^2$ and $10\,\text{GeV}<\nu<32\,\text{GeV}$. 
The dashed blue vertical lines enclose the region accepted for analysis. Here, Monte Carlo also includes $\pi^0$ background.
}
\end{figure}

Exemplary results of this comparison are displayed in Fig.~\ref{fig:ex_vars} using two variables that characterize the kinematics of the recoiling target particle.
Figure~\ref{fig:ex_vars}(a) shows the difference between the measured and the predicted azimuthal angle, $\Delta \Phi$, and Fig.~\ref{fig:ex_vars}(b) the difference between the measured and the predicted transverse momentum, $\Delta p_{\text{T}}$.
Here, $\Phi$ and $p_{\text{T}}$ are given in the laboratory system.

Figure~\ref{fig:ex_vars} shows additionally a comparison between the data and
the sum of Monte Carlo yields
that includes all single-photon production mechanisms, i.e. BH, DVCS and their interference, as well as the $\pi^0$ background estimates. The Monte Carlo simulations for all these 
mechanisms are based on the HEPGEN generator~\cite{HEPGEN,HEPGEN_pp}. The adopted DVCS amplitude follows the model of Refs.~\cite{FFS1,FFS2}, which was originally proposed to describe the DVCS data measured at very small $\xBj$ at HERA, with modifications required for COMPASS (see Refs.~\cite{HEPGEN,Philipp} and references therein). For the BH amplitude and the interference term, the formalism of Ref.~\cite{mueller2002} is used replacing the approximate expressions for the lepton propagators P1 and P2 by the exact formulae that take into account the non-zero mass of the lepton. The HEPGEN simulations are normalized to the total integrated luminosity of the data. The simulations are also used for the calculation of the spectrometer acceptance. 

In order to identify background events originating from $\pi^0$ production, where one photon of the $\pi^0$ decay is detected in an electromagnetic calorimeter but falls short of the above given threshold, the single-photon candidate is combined with every neutral cluster below threshold. The event is excluded if a $\pi^0$ with $|m_{\gamma\gamma}-m_{\pi^0}^{PDG}|<20\,\text{MeV}/c^2$ can be reconstructed. 
This corresponds to about 1.5 standard deviations of the mass resolution. The number of excluded events is used below to normalise the $\pi^0$ Monte Carlo simulation.

Background originating from $\pi^0$ production, where one photon of the $\pi^0$ decay remains undetected, is estimated using a Monte Carlo simulation that is normalised to the aforementioned excluded fraction of $\pi^0$ events. 
This simulation
, which is denoted as $\pi^0$ background in Fig.\ref{fig:ex_vars}, is the sum of two components. First, the HEPGEN generator 
uses the parameterisation of Ref.~\cite{GK11} for the cross section of the exclusive reaction $\mu p \rightarrow \mu p \pi^0$. Secondly, the LEPTO~6.5.1 generator with the COMPASS high-$p_{\text{T}}$ tuning \cite{highpt}
is used to simulate the tail of non-exclusive $\pi^0$ production, which is accepted by our experimental selections.
Comparing the two components to the data allows the determination of their relative normalisation.

\begin{figure}[h!] 
\begin{center}  
  \includegraphics[width=1.0\linewidth]{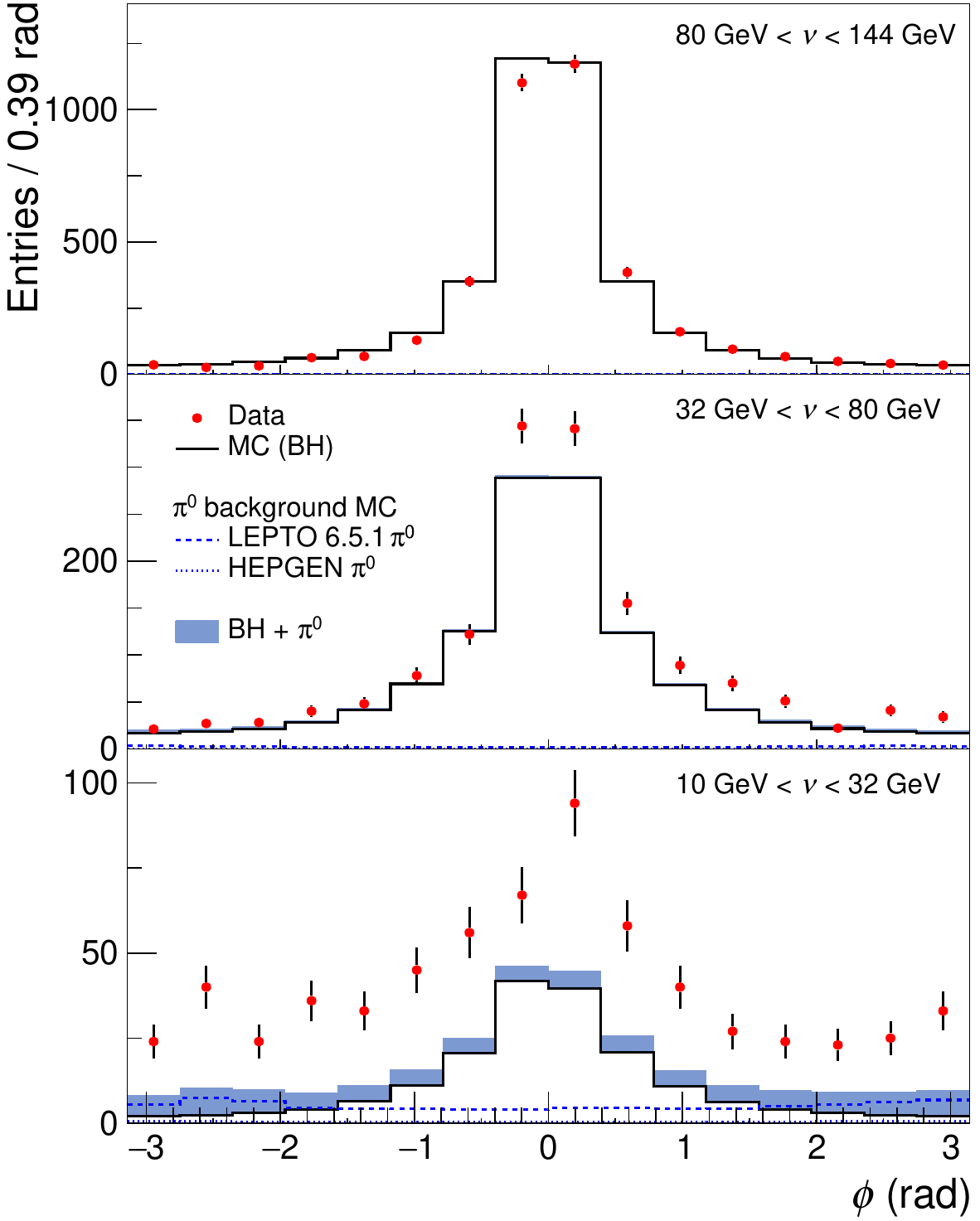}
\end{center} 
\vspace{-0.25cm}
  \caption{\label{fig:3phi}
  Number of reconstructed single-photon events as a function of $\phigg$ in three regions of $\nu$ for $1\,(\text{GeV/}c)^2<Q^2<5\,(\text{GeV/}c)^2$ and $0.08\,(\text{GeV/}c)^2<|t|<0.64\,(\text{GeV/}c)^2$.
  Error bars represent statistical uncertainties.
  Additionally shown are the sum of a Monte Carlo simulation of the BH process and the two components of the $\pi^0$ contamination described in the text. Note that the yield of the HEPGEN $\pi^0$ contribution is very small and at most
  0.01, 0.2 or 0.6 entries per $\phigg$-bin in the panels from top to bottom, respectively. 
  The data in this figure are not yet corrected for $\pi^0$ background.}
\end{figure}

After the application of the above described selection criteria a kinematic fit is performed, which is constrained by requiring a single-photon final state in order to obtain the best possible determination of all kinematic parameters in a given event. Figure~\ref{fig:3phi} shows the number of selected single-photon events as a function of $\phigg$ for three different regions in the virtual-photon energy $\nu$.
The data are compared to the sum of a Monte Carlo simulation of the BH process only, which is normalised to the total integrated luminosity of the data, and the estimated $\pi^0$ contamination. For large values of $\nu$, the data agree reasonably well with the expectation that only the BH process contributes. For intermediate and small values of $\nu$, sizable contributions from the DVCS process and the BH-DVCS interference are observed.

From here on, the analysis is performed in the region of small $\nu$ using a three-dimensional equidistant grid with four bins in $|t|$ from 0.08\,\qsqU \ to 0.64\,\qsqU \,, 11 bins in $\nu$ from 10\,GeV to 32\,GeV, and four bins in $Q^2$ from 1\,\qsqU \ to 5\,\qsqU. 
For each bin the acceptance correction is applied and the contribution of the BH process is subtracted together with the estimated $\pi^0$ contamination. The photon flux factor is applied on an event-by-event basis according to Eq.~(\ref{eq:flux_relation}). In every of the four bins in $|t|$, the mean value of the cross section is obtained by averaging over $Q^2$ and $\nu$. 
When determining the cross section in bins of $\phigg$, no significant dependence on 
$\phigg$ is observed. According to Eq.~(\ref{eq:phi_mods}), the extracted result is in such a case sensitive to the quantity $c_0^{DVCS}$ only.

\section{Results}
\label{sec:results}

\begin{figure}[h!] 
\begin{center}  
  \includegraphics[width=1.0\linewidth]{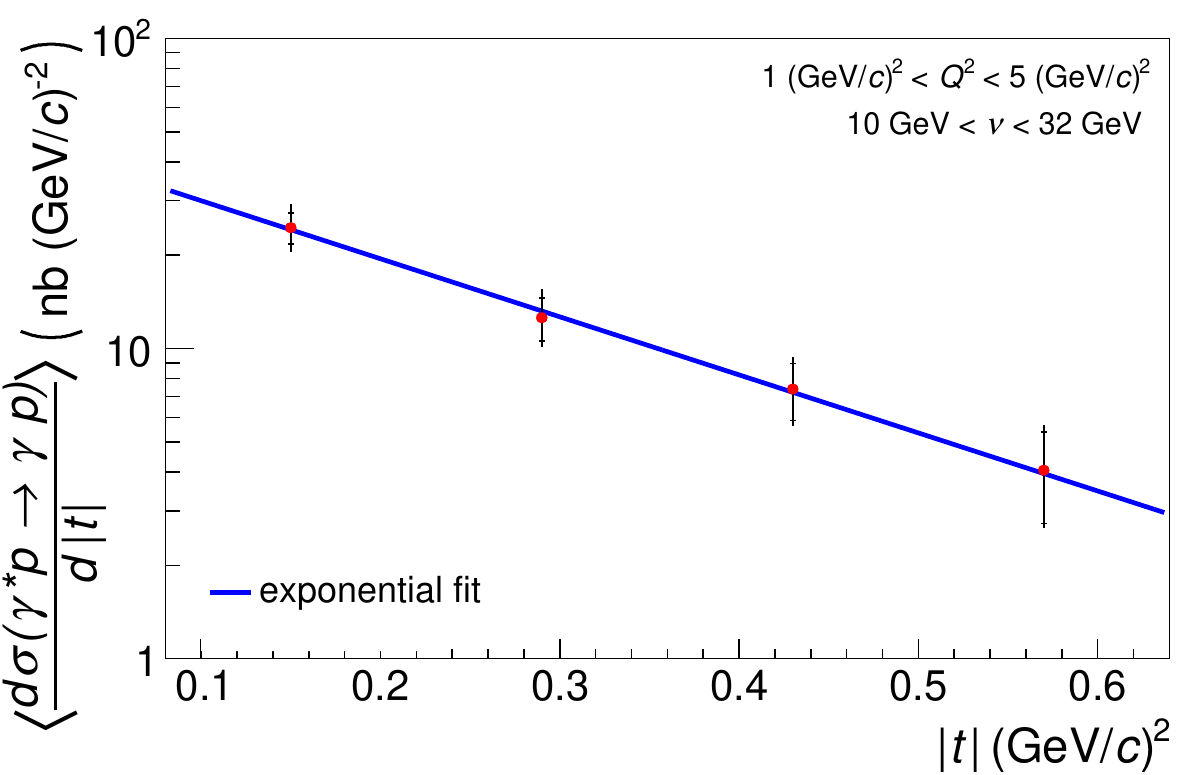}
\end{center}
\vspace{-0.25cm}
  \caption{\label{fig:x_sec}
  Differential DVCS cross section as a function of $|t|$. The mean value of the cross section is shown
  at the center of each of the four $|t|$-bins. The blue curve is the result of a binned maximum likelihood fit of an exponential function to the data. This fit integrates the exponential model over the respective t-bins and does not use their central values, which are used for illustration only. The probability to observe a similar or better agreement of the data with the blue curve is approximately $7$\%. Here and in the next figure, inner error bars represent statistical uncertainties and outer error bars the quadratic sum of statistical and systematic
  uncertainties. }
\end{figure}

The $t$-dependence of the 
extracted $\mu^+$ and $\mu^-$ cross section average is shown in Fig.~\ref{fig:x_sec}, with the numerical values given in Tab.~\ref{tab:1}.
\begin{table}[h!] 
\caption{\label{tab:1} Values of the extracted DVCS cross section: The quantity $\langle \frac{\di\sigma}{d|t|} \rangle$ denotes the average of the measured differential $\mu^+$ and $\mu^-$ DVCS cross sections in the indicated $|t|$-bin. Apart from the integration over $t$, the cross section is integrated over $Q^2$ and $\nu$ and divided by the product of the respective bin widths, as  indicated in Fig.~\ref{fig:x_sec}. In addition, the mean values for $Q^2$ and $\nu$ are given for each of the four bins. These mean values are weighted averages with the weight being the virtual-photon proton 
cross section.}
\begin{center}
\begin{tabular}{ l c c c }
\toprule
 $\frac{|t|-\text{bin}}{(\text{GeV}/c)^2}$ \hspace{5pt}  &  $\frac{\langle \frac{\di \sigma}{d|t|} \rangle}{\text{nb}(\text{GeV}/c)^{-2}}$ & $\frac{\langle Q^2 \rangle}{(\text{GeV}/c)^2}$ & $\frac{\langle \nu \rangle}{\text{GeV}}$  \\  
 \midrule
 &  \\[-6pt]
[0.08, 0.22] \hspace{5pt} & $24.5 \pm 2.8_{\text{stat}} \ {}^{+3.7}_{-2.9}\big\rvert_{\text{sys}}$ & 1.79 & 19.5   \\[4pt]  
[0.22, 0.36] \hspace{5pt} & $12.6 \pm 2.0_{\text{stat}} \ {}^{+2.2}_{-1.5}\big\rvert_{\text{sys}}$ & 1.77 & 18.8  \\[4pt]  
[0.36, 0.50] \hspace{5pt} & $\textcolor{white}{0}7.4 \pm 1.6_{\text{stat}} \ {}^{+1.3}_{-0.9}\big\rvert_{\text{sys}}$ & 1.91 & 18.6  \\[4pt]  
[0.50, 0.64] \hspace{5pt} & $\textcolor{white}{0}4.1 \pm 1.3_{\text{stat}} \ {}^{+1.0}_{-0.5}\big\rvert_{\text{sys}}$ & 1.77 & 20.1  \\
\bottomrule
\end{tabular}
\end{center}
\end{table}
The observed $t$-dependence of the DVCS cross section can be well described by a single-exponential function $e^{-B |t|}$.
The four data points are fitted using a binned maximum-likelihood method, 
where the weights take into account all corrections mentioned above.
The result on the $t$-slope,
\begin{equation} 
\label{eq:B-result}
B=(4.3 \ \pm \ 0.6_{\text{stat}}\ _{- \ 0.3}^{+ \ 0.1}\big\rvert_{\text{sys}})\,\,(\text{GeV}/c)^{-2},
\end{equation}
is obtained at the average kinematics $\langle W \rangle = 5.8\,\text{GeV/}c^2$, $\langle Q^2 \rangle = 1.8\,(\text{GeV/}c)^2$ and $\langle \xBj \rangle = 0.056$. 

In Tab.~\ref{tab:sys_t}, the important contributions to the systematic uncertainties on the values of cross sections and exponential slope are shown, arranged in three groups. The first group contains symmetric contributions due to uncertainties in the determination of the beam
flux, possible variations of the energy and momentum balance in the kinematic fit and the statistical uncertainty of the background subtraction. The second group contains systematic uncertainties related to corrections that were applied to the measured cross section.
The subtracted amount of $\pi^0$ background is translated into an uni-directional systematic uncertainty of up to $+12$\%, which is related to the detection of photons and originates from a possible bias on the low energy-thresholds of the electromagnetic calorimeters. 
As radiative corrections to the DVCS process are model dependent, they are not applied but instead also included as an uni-directional systematic uncertainty.
The third group contains the largest contribution to the systematic uncertainty.
It is linked to the normalisation of the data in the large $\nu$-range with respect to the
Bethe-Heitler contribution, when comparing data taking with positively and negatively charged muon beam.
It is asymmetric and amounts to at most $+19$\% and $-9$\%
for large values of $|t|$. The total systematic uncertainty $\Sigma$ is obtained as quadratic sum of 
all components shown in Tab.~\ref{tab:sys_t}.

The main systematic uncertainty on the slope $B$
is uni-directional with a value of $-5$\% and originates from the normalisation of the $\pi^0$ background. Note that the systematic uncertainties of the four data points for the cross section are strongly correlated, so that for the slope value a considerably smaller systematic uncertainty is obtained. More details on systematic uncertainties are given in Ref.~\cite{Philipp}.

\begin{table}[hbt]
\centering

\caption{Columns 1 and 2 show the relative systematic uncertainties on the measured cross section in bins of $|t|$, columns 3 and 4 show those on the fitted slope of the cross section. All values are given in percent. Note that the uni-directional systematic uncertainty $\sigma_{\uparrow}$ ($\sigma_{\downarrow}$) has to be used with positive (negative) sign.}
\label{tab:sys_t}

\vspace{0.5cm}
\begin{tabular}{l c c c c}
\toprule
Source                  & $\sigma^t_{\uparrow}$ &   $\sigma^t_{\downarrow}$ &   $\sigma^B_{\uparrow}$ &     $\sigma^B_{\downarrow}$\\
\midrule
muon flux               &          3&3&&\\
kinematic fit           &             3&3&0&0\\
background stat. unc.   &             2 - 5&2 - 5&2&2\\ \midrule
background norm.        &            0&6 - 12&0&5\\
radiative corr.         &            0&4 - 6&0&1\\ \midrule
reconstr. unc.          &    13 - 19& 9 &0&2\\
\midrule
$\Sigma$ &               15 - 23 & 11 - 12 &2&6\\
\bottomrule
\end{tabular}

\end{table}

\section{Interpretation}
\label{sec:interpretation}

\begin{figure}[h!]
\begin{center}  
  \includegraphics[width=1.0\linewidth]{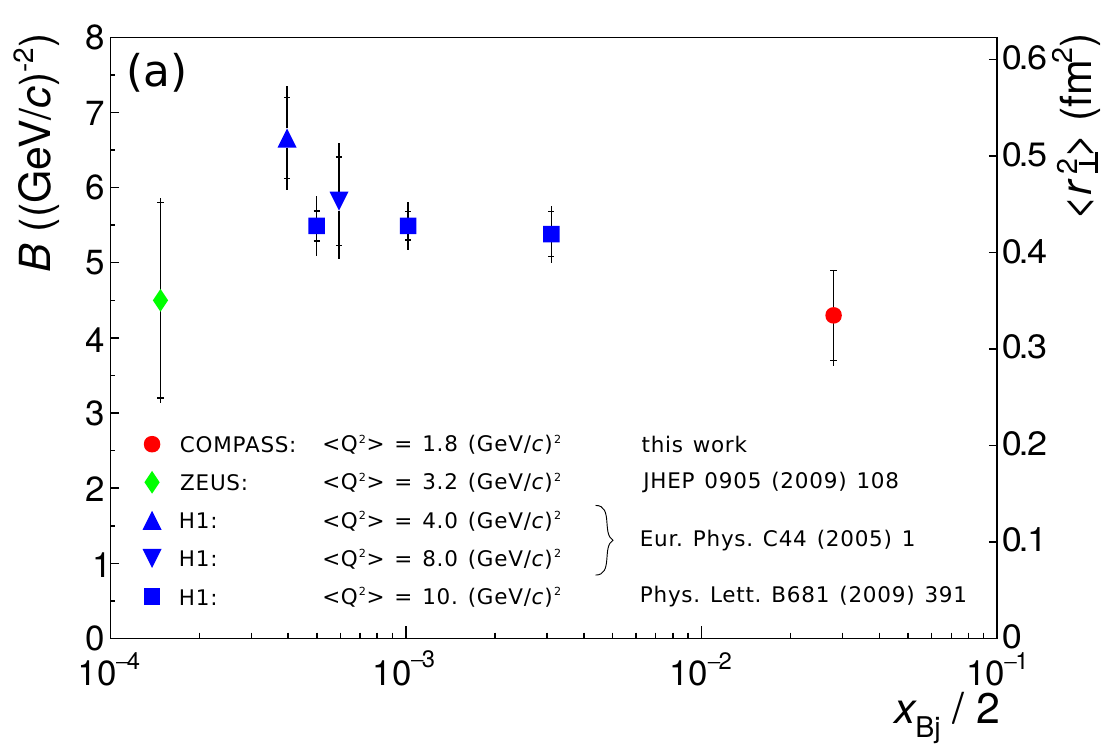}
  
  \includegraphics[width=1.0\linewidth]{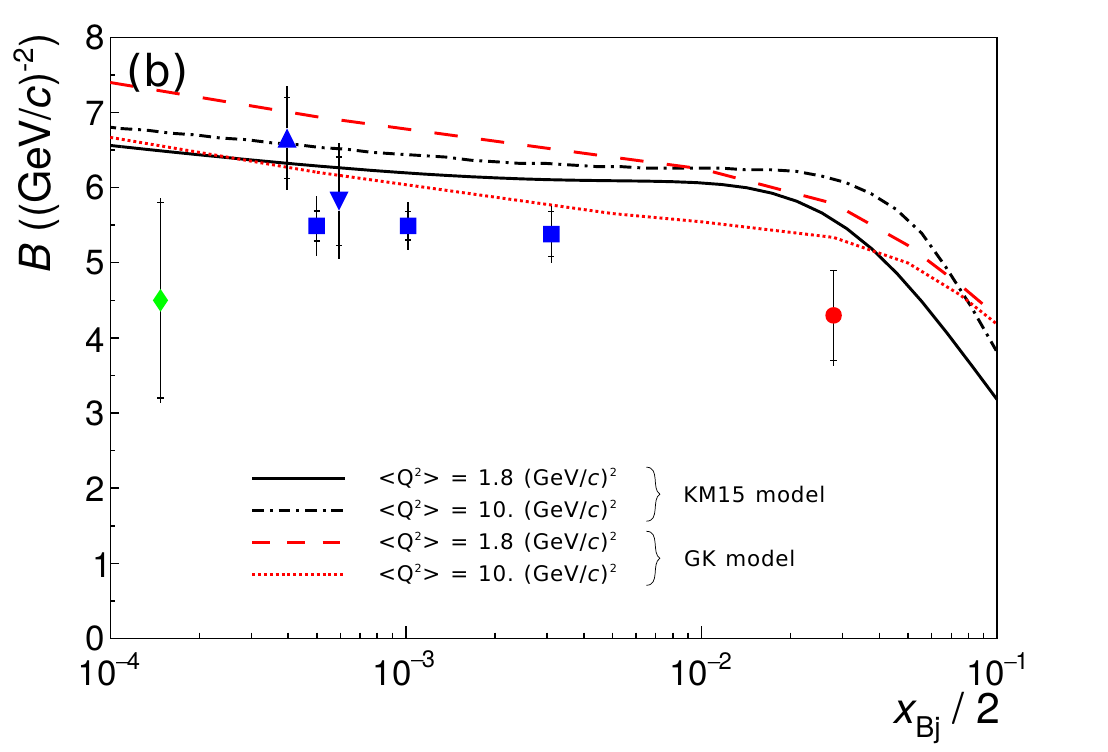}
  \end{center}
\vspace{-0.25cm}
  \caption{\label{fig:b_comp} (a) Results from COMPASS and previous measurements by H1 \cite{BMeas:H1:1,BMeas:H1:2} and ZEUS \cite{BMeas:ZEUS} on the $t$-slope parameter $B$, or equivalently the average squared transverse extension of partons in the proton, $\langle r_{\perp}^2 \rangle$, as probed by DVCS at the proton longitudinal momentum fraction $\xBj /2$ (see text). Inner error bars represent statistical and outer ones the quadratic sum of statistical and systematic uncertainties. (b) Same results compared to the predictions of the GK \cite{GK1,GK2,GK3} and KM15 \cite{Kumericki:2009uq,Kumericki:2015lhb} models.}

\end{figure}

This Letter presents the first measurement of the $|t|$-dependence of the differential DVCS cross section in the intermediate $\xBj$-region, which can be described by a single-exponential function $e^{-B |t|}$. 
Using Eq.~(\ref{eq:radius}), the fitted slope $B$ of the measured 
$|t|$-dependence of the DVCS cross section is converted into the transverse 
extension of partons in the proton, as probed by DVCS at about $\langle \xBj \rangle /2 = 0.028$: 

\begin{equation} 
\label{eq:r-result}
\sqrt{\langle r_{\perp}^2 \rangle} = (0.58 \ \pm \ 0.04_{\text{stat}}\ _{- \ 0.02}^{+ \ 0.01}\big\rvert_{\text{sys}} \ 
\pm 0.04_{\text{model}}
)\, \text{fm}.
\end{equation}
The determination of the model uncertainty is explained below.
Figure~\ref{fig:b_comp} (a) shows our result together with those obtained by earlier high-energy experiments that used the same method to determine the DVCS cross section and extract the $t$-slope parameter $B$, or equivalently  the average squared transverse extension of partons in the proton, $\langle r_{\perp}^2 \rangle$. 
We note that the results of the HERA collider experiments H1~\cite{BMeas:H1:1,BMeas:H1:2} and ZEUS~\cite{BMeas:ZEUS} were obtained at higher values of $Q^2$ as compared to that of the COMPASS measurement. Also, while our measurement 
probes the transverse extension of partons in the proton in the intermediate $\xBj$ range, the measurements at HERA are sensitive to values of $\xBj/2$ below $10^{-2}$. 

As described e.g. in Ref.~\cite{Moutarde:2018kwr}, the slope $B$ of the $|t|$-dependence of the DVCS cross section can be converted into the transverse extension of partons in the proton assuming i) the dominance of the imaginary part of the CFF $\mathcal{H}$, and ii)~a negligible  effect of a non-zero value of the skewness $\xi \approx \xBj/2$ in the actual measurement. Both assumptions are expected to hold at small values of $\xBj$.

In the following, we interpret our measurement of the $B$-slope at leading order in $\alpha_S$ and at leading twist.
In such a case, the spin-independent DVCS cross section is only sensitive to the quantity $c_0^{DVCS}$ that is related at  small $\xBj$ to the CFFs $\mathcal{H}$, $\mathcal{\tilde{H}}$ and $\mathcal{E}$ as~\cite{mueller2002}:
\begin{equation}
c_0^{DVCS} 
\propto
 4(\mathcal{H}\mathcal{H^*} + \mathcal{\tilde{H}}\mathcal{\tilde{H}^*})
+  \frac{t}{M^2} \mathcal{E}\mathcal{E^*}.
\end{equation}
In the  $\xBj$-domain of COMPASS, $c_0^{DVCS}$ is dominated by the imaginary part of the CFF $\mathcal{H}$. 
In this region, the contributions by the real part of $\mathcal{H}$ and by other CFFs
amount to about 3\% when calculated using the GK model~\cite{GK1,GK2,GK3}
ported to the PARTONS  framework~\cite{Berthou:2015oaw} and to about 6\% when using the KM15 model~\cite{Kumericki:2009uq,Kumericki:2015lhb}. Using the second value, the systematic model uncertainty related to assumption i) above is estimated to be about $\pm 0.03\,\text{fm}$.

The skewness $\xi$ is equal to one half of the
longitudinal momentum fraction transferred between the initial and final proton.
A strict relation between the slope $B$ and $\langle r_{\perp}^2 \rangle$ only exists for $\xi = 0.$
A non-zero value of $\xi$ introduces an additional uncertainty on $\langle r_{\perp}^2 \rangle$ that is related to a shift of the center of the reference system,
in which $\langle r_{\perp}^2 \rangle$ is defined~\cite{Diehl:2002he}.
Using the GK model,
we estimate the corresponding systematic uncertainty 
regarding assumption ii) above 
to be about $\pm 0.02\,\text{fm}$. The value for the model uncertainty given in Eq.~\ref{eq:r-result} is obtained by quadratic summation of the two components.

The same data as presented in Fig.~\ref{fig:b_comp} (a) are shown in  Fig.~\ref{fig:b_comp} (b), compared to calculations of the phenomenological GK and KM15 models, which describe
the data reasonably well in the low and medium $\xBj$ range. Even taking into account the relatively small effect of $Q^2$ evolution, some scale offset between data and models seems to exist. When comparing our result on the transverse extension of partons in the proton to the lowest-$Q^2$ result of H1, there is an indication for shrinkage, i.e. a decrease of the B-slope with $x_{Bj}$, 
at the level of about 2.5 standard deviations of the combined uncertainty.

In order to reliably determine the full $\xBj$-dependence of the transverse extension of partons in the proton,
a global phenomenological analysis 
using all results from DVCS experiments at HERA, CERN, and JLab appears necessary to pin down the imaginary part of CFF $\mathcal{H}$, and eventually the GPD $H$ itself.
At leading order in $\alpha_s$ and at leading twist, such analyses~\cite{Kumericki:2013br,Kumericki:2015lhb,Kumericki:2016ehc,Dupre:2016mai,Dupre:2017hfs,Moutarde:2018kwr} have already been performed in order to interpret the 
results of those experiments that access the high-$\xBj$ region, i.e. mostly the valence-quark sector probed by  HERMES and at JLab (see e.g. Ref.~\cite{Moutarde:2018kwr} for a list of experimental results).
In such a global analysis, the $Q^{2}$ evolution and all necessary corrections have to be included that are required under the kinematic conditions of the respective experiments. 
Possibly, also results on exclusive-meson production may be included. Eventually, this may allow one to disentangle the contributions of the different parton species to the transverse size of the proton as a function of the average longitudinal momentum fraction carried by its constituents.

\section{Summary}

In summary, using exclusive single-photon muoproduction we have measured the $t$-slope of the deeply virtual Compton scattering cross section at $\langle W \rangle = 5.8\,(\text{GeV/}c)^2$, $\langle Q^2 \rangle = 1.8\,(\text{GeV/}c)^2$ and $\langle \xBj \rangle = 0.056$, which leads to the slope value 
$B=(4.3 \ \pm \ 0.6_{\text{stat}}\ _{- \ 0.3}^{+ \ 0.1}\big\rvert_{\text{sys}})\,(\text{GeV}/c)^{-2}$. For an average longitudinal momentum fraction carried by the partons in the proton of about $\langle \xBj \rangle /2 =  0.028$, we find a transverse extension of partons in the proton of $\sqrt{\langle r_{\perp}^2 \rangle} = (0.58 \ \pm \ 0.04_{\text{stat}}\ _{- \ 0.02}^{+ \ 0.01}\big\rvert_{\text{sys}}\ \pm \ 0.04_{\text{model}})\, \text{fm}$.

\section*{Acknowledgements}
We gratefully acknowledge the support of the CERN
management and staff, as well as the skill and effort of the technicians of our collaborating institutes. We thank Pierre Guichon for the evaluation of the Bethe-Heitler contribution taking into account the muon mass, Sergey Golos\-kokov and Peter Kroll for their continuous support with model predictions for the $\pi^0$ background extraction, as well as Kresimir Kumericki and Dieter Mueller for providing their theoretical predictions on the t-slope parameter B.

\end{document}